# One-dimensional energy spectra in three-dimensional incompressible homogeneous isotropic turbulence [†]

RAN Zheng[1*], WANG Yao-yao [1], YUAN Xing-jie [1]

*1 Shanghai Institute of Applied Mathematics and Mechanics, Shanghai University, Shanghai 200072, P.R.China*

The paper investigates the detailed features of one-dimensional energy spectra in three-dimensional isotropic turbulence, based on the exact solution of Karman-Howarth equation. Particular interest will be paid on the degree to which spectral scaling can lead the spectral data to be collapsed. The theory appears to be consistent with the wealth of experimental data (G.Comte-Bellot and S. Corrsin, 1971) at least in low Taylor microscale Reynolds number.

# 1 Introduction

In 1935 G.I.Taylor [1] introduced the notion of statistical homogeneity and isotropy: a step which took turbulence theory into the realm of physics, rather than engineering. In a subsequent paper [2], the introduction of the energy spectrum in wave number virtually completes this process and, as we see, the calculation of this spectrum provides a major goal for fundamental turbulence theory. Th.v. Karman and L.Howarth [3] in 1938 developed this theory still further. The double and triple-correlation functions were introduced, and a differential equation connecting them was established. This differential equation is of fundamental importance in the statistical theory of turbulence, and much effort has been spent in order to extract as much knowledge from it as possible. Karman-Howarth equation should occupy an important place in any practical theories of turbulence [4].

It is well known that : Beside models based on the solution of the Karman-Howarth equation, there are other approaches to theoretical investigation of isotropic turbulence dynamical such as analysis of possible similarity solutions of nonclosed Karman-Howarth equation or its spectral analog; attempts to close equation with the use of an energy spectrum [5-11]. In 1944, Sedov [12], proposed another way of Karman-Howarth equation closure. Some analytical results based on this way of closure were in good agreement with experimental data [13, 14]. Recently, the existence new Sedov-type solution has been found and also gives some new insight on the behaviour of isotropic turbulence [15, 16, 17, 18], especially for the turbulence energy spectra.

The paper will investigate the detailed features of one-dimensional velocity spectra in three-dimensional isotropic turbulence. Particular interest will be paid on the new one-dimensional energy spectra based on the exact solution of Karman-Howarth equation. An example is presented and used to evaluate the relative merits of the new theory.

This paper is organized as follows. In section 2 we will first go over the model describing. In the following sections we present the new formula and the calculated results and comparisons with observational data of the one-dimensional energy spectra of isotropic turbulence. Finally section 6 is the conclusion.

## 2 Theoretical background

We will consider isotropic turbulence governed by the incompressible Navier-Stokes equations. The two-point double and triple longitudinal velocity correlation denoted by $f(r,t)$ and $h(r,t)$ respectively, are defined in a standard way. For isotropic turbulence, they satisfy the Karman-Howarth equation

$$\frac{\partial}{\partial t}(bf) + 2b^{\frac{3}{2}}\left(\frac{\partial h}{\partial r} + \frac{4h}{r}\right) = 2\nu b\left(\frac{\partial^2 f}{\partial r^2} + \frac{4}{r}\frac{\partial f}{\partial r}\right) \tag{1}$$

where $(r,t)$ is the spatial and time coordinates, $\nu$ is the kinematics viscosity, and $b = \overline{u^2}$ denotes the turbulence intensity.

However, these equations are not sufficient for the description of the time evolution of isotropic turbulence. Many workers have therefore tried to augment Karman-Howarth equation by special hypotheses containing additional information about the time variation of the correlation and spectral functions. Self-preservation hypotheses are such thing-which imposes certain restrictions on the general character of the variation of the correlation and spectral functions with time. Let us suppose that the functions $f(r,t)$ and $h(r,t)$ preserve the same form as time increases with only the scale varying. Such functions will be termed as "self-preserving". Following von Karman and Howarth, we introduce the new variables

$$\xi = \frac{r}{l(t)} \tag{2}$$

Where $l = l(t)$ is a uniquely specified similarity length scale. For an isotropic turbulence to be self-preserving in the sense of von Karman and Howarth [3] and Batchelor [4], we must have

$$f(r,t) = f(\xi) \tag{3}$$

$$h(r,t) = h(\xi) \tag{4}$$

For complete self-preservation, all scales of the turbulence, namely, those in the full range of $0 \leq r \leq \infty$ must decay according to (3) and (4). We will focus our attention on complete self-preserving solutions in the analysis below. For complete self-preserving isotropic turbulence, Sedov [12, 13] found an ingenious way to obtain two equations from one. The main results could

be listed as following:

[1]The equation of the second order longitudinal correlation coefficient can be written in the form

$$\frac{d^2 f}{d\xi^2} + \left(\frac{4}{\xi} + \frac{a_1}{2}\xi\right)\frac{df}{d\xi} + \frac{a_2}{2} f = 0 \tag{5}$$

With boundary conditions

$$f(0) = 1. \tag{6}$$

$$f(\infty) = 0. \tag{7}$$

And $a_1, a_2$ are constant coefficients.

[2]The corresponding scaling (length and energy) equations read:

$$\frac{1}{\sqrt{b}}\frac{dl}{dt} = a_1\left(\frac{\nu}{\sqrt{bl}}\right) + p \tag{8}$$

$$-\frac{l}{b^{\frac{3}{2}}}\frac{db}{dt} = a_2\left(\frac{\nu}{\sqrt{bl}}\right) \tag{9}$$

Where $p$ are constants of integration. These are the scaling equations of isotropic turbulence.

[3]Substituting leads the equation determing the third correlation coefficient

$$\frac{dh}{d\xi} + \frac{4}{\xi}h = \frac{p}{2}\xi\frac{df}{d\xi} \tag{10}$$

These equations contain three arbitrary constants $a_1, a_2, p$. By putting $a_1 = \frac{1}{4}$, $a_2 = 10\alpha > 0$ and $\frac{r^2}{l^2} = \xi^2 = \varsigma$, Sedov obtained the same solution for $f(\varsigma)$ as in the small fluctuations case, namely,

$$f(\varsigma) = {}_1F_1\left(10\alpha, \frac{5}{2}, -\frac{\varsigma}{8}\right) \tag{11}$$

Where ${}_1F_1(\alpha, \gamma, z)$ is the confluent hypergeometric function. Meanwhile, Sedov obtained

$$h = \frac{2p}{4\alpha - 1}[f'(\xi) + \alpha\xi f(\xi)] \tag{12}$$

Sedov [12, 13] also gave the solution to determine $l$ and $b$ as functions of time.

Nonetheless, it appears that the Sedov's method has never been reinvestigated in depth subsequent to the earlier work after 1944. Further investigation on this subject could be seen in Refs. [15-18]. It is believed that Sedov's closure method has many interesting features that has not yet been fully understood and is worthy go further study.

# 3 One dimensional energy spectra

In the following analysis, unlike in the previous studies, we introduce two alternative parameters denoted by $a_1, \sigma$, while the Sedov-type solution could be rewritten as

$$f(\xi) = {}_1F_1\left(\sigma, \frac{5}{2}, -\frac{a_1}{4}\xi^2\right) \tag{13}$$

One-dimensional energy spectra could be deduced directly from this solution.

The one-dimensional spectra that are most often measured are the one-dimensional Fourier transforms of a longitudinal correlation. There is no uniformity of notation for one-dimensional energy spectra. [5, 6, 7]. We shall adopt the same convention as Tennekes and Lumley for the one-dimensional spectra of $u^2 f$, that is $F_{11}$ [7].

The longitudinal spectrum appears particularly in experimental papers as the quantities most commonly measured in experiment. It is

$$F_{11}(k,t) = \frac{1}{\pi} \int_0^\infty u^2 f(r,t) \cos(kr) dr \tag{14}$$

With inverse transform,

$$u^2 f(r,t) = 2 \int_0^\infty F_{11}(k,t) \cos(kr) dk \tag{15}$$

Here $f$ is the usual longitudinal correlation function. Of course, $F_{11}(k,t)$ are simply the one-dimensional energy spectra of $u_x(x,0,0)$.

By using the integral formula:

$$\int_0^\infty {}_1F_1(a;c;-t^2) \cos(2zt) dt = \sqrt{\frac{\pi}{2}} \frac{\Gamma(c)}{\Gamma(a)} z^{2a-1} e^{-z^2} U\left(c - \frac{1}{2}, a + \frac{1}{2}, z^2\right) \tag{16}$$

Where ${}_1F_1(a,c,z)$ is Kummer's hypergemometric function, and $U(a,c,z)$ is another function closely related to Kummer's function defined by

$$U(a,c,z) = \frac{\Gamma(1-c)}{\Gamma(1+a-c)} {}_1F_1(a,c,z) + \frac{\Gamma(c-1)}{\Gamma(a)} z^{1-c} {}_1F_1(1+a-c, 2-c, z) \tag{17}$$

Where $\Gamma(z)$ is the usual Gamma function. For the Sedov-type solution, according to formula (16),

for isotropic turbulence, we have

$$F_{11}(k,t) = \frac{b}{\pi} \cdot \frac{l}{\sqrt{a_1}} \cdot \sqrt{\pi} \cdot \frac{\Gamma\left(\frac{5}{2}\right)}{\Gamma(\sigma)} \cdot \left(\frac{l}{\sqrt{a_1}} k\right)^{2\sigma-1} \cdot e^{-\frac{l^2}{a_1}k^2} \cdot U\left(2, \sigma+\frac{1}{2}, \frac{l^2}{a_1}k^2\right) \quad (18)$$

This is the exact result of the longitudinal spectra based on the Sedov-type solution.

## 4 The spectral scaling

Before carrying out the examination of the spectral data, however, it is useful to review what comparison will be made and what kind of agreement (or disagreement) with the theory might be expected. Since the data are usually presented as a function of position behind the grid, it is interesting to note how the various scaling parameters vary with the position behind the grid. Four types of scaling are of interest: the integral length scaling and the Taylor scaling proposed here, the von Karman-Howarth scaling, and the Kolmogorov scaling. The proposed Kolmogorov variable scaling consistence with the measure spectral data.

[1]The integral scaling:

We set

$$z = \frac{l^2}{a_1} \cdot k^2, \quad (34)$$

In view of the identity

$$z = \frac{1}{a_1} \cdot \left[\frac{l}{L}\right]^2 \cdot [Lk]^2, \quad (35)$$

from which it can be concluded that

$$z = [\Phi(\sigma)]^{-1} \cdot [Lk]^2, \quad (36)$$

Let

$$y_L \equiv F_{11}(k,t) \cdot [bL]^{-1}, \quad (37)$$

It can be see that

$$y_L = \frac{1}{\sqrt{\pi}} \cdot \frac{\Gamma\left(\frac{5}{2}\right)}{\Gamma(\sigma)} \cdot \frac{1}{\Phi(\sigma)} \cdot (z)^{\sigma-\frac{1}{2}} \cdot e^{-z} \cdot U\left(2, \sigma+\frac{1}{2}, z\right) \quad (38)$$

[2] Taylor micro scale scaling

In the same time, in view of the identity

$$z = \frac{1}{a_1} \cdot \left[\frac{l}{\lambda}\right]^2 \cdot [\lambda k]^2, \tag{39}$$

From which it can be concluded that

$$z = \frac{1}{5} \cdot \sigma \cdot [\lambda k]^2, \tag{40}$$

Let

$$y_T \equiv F_{11}(k,t) \cdot [b\lambda]^{-1}, \tag{41}$$

It can be see that

$$y_T = \sqrt{\frac{\sigma}{5\pi}} \cdot \frac{\Gamma\left(\frac{5}{2}\right)}{\Gamma(\sigma)} \cdot z^{\sigma - \frac{1}{2}} \cdot e^{-z} \cdot U\left(2, \sigma + \frac{1}{2}, z\right), \tag{42}$$

[3] Karman-Howarth scaling:

For the third suggested scaling, in view of the identity

$$z = \frac{1}{a_1} \cdot [lk]^2, \tag{43}$$

Let

$$y_V \equiv F_{11}(k,t) \cdot [bl]^{-1}, \tag{44}$$

$$y_V = \frac{1}{\sqrt{\pi a_1}} \cdot \frac{\Gamma\left(\frac{5}{2}\right)}{\Gamma(\sigma)} \cdot z^{\sigma - \frac{1}{2}} \cdot e^{-z} \cdot U\left(2, \sigma + \frac{1}{2}, z\right) \tag{45}$$

[4] Kolmogorov scaling:

Given the two parameters $\varepsilon$ (the rate of turbulent dissipation) and $\nu$, there are unique length, velocity, and time scales that can be formed. These are the Kolmogorov scales:

$$\eta = [\nu^3/\varepsilon]^{\frac{1}{4}} \tag{46}$$

$$u_\eta = [\varepsilon\nu]^{\frac{1}{4}} \tag{47}$$

$$\tau_\eta = [\nu/\varepsilon]^{\frac{1}{2}} \tag{48}$$

Having identified the Kolmogorov scales, we can now state a consequence of the hypotheses that demonstrates their potency, and clarifies the meaning of the phrases: similarity hypothesis and universal form.

The turbulent kinetic energy

$$K \equiv \frac{1}{2}\overline{u_i u_i} = \frac{3}{2}b \qquad (49)$$

Which is the solution of the differential equation?

$$\frac{dK}{dt} = -\varepsilon \qquad (50)$$

Hence, for any complete self-preservation isotropic turbulence, we must have

$$\varepsilon = \frac{3}{2} \cdot \frac{a_2 \nu b}{l^2} \qquad (51)$$

It follows from (19) that

$$\begin{aligned}
\eta &= \left[\frac{\nu^3}{\varepsilon}\right]^{\frac{1}{4}} \\
&= \nu^{\frac{3}{4}} \cdot \left[\frac{3}{2} \cdot \frac{a_2 \nu b}{l^2}\right]^{-\frac{1}{4}} \\
&= \left[\frac{2}{3}\right]^{\frac{1}{4}} \cdot [a_2]^{-\frac{1}{4}} \cdot [\nu]^{\frac{1}{2}} \cdot [b]^{-\frac{1}{4}} \cdot [l]^{\frac{1}{2}}
\end{aligned} \qquad (52)$$

In view of the identity

$$z = \frac{1}{a_1} \cdot \left[\frac{l}{\eta}\right]^2 \cdot [\eta k]^2, \qquad (53)$$

From which it can be concluded that

$$z = 2\sigma \cdot \left[\frac{3}{20}\right]^{\frac{1}{2}} \cdot [R_\lambda] \cdot [\eta k]^2, \qquad (54)$$

Which is the turbulent Reyonlds number based on the Taylor micro scale?

Let

$$y \equiv F_{11}(k,t) \cdot \left[\varepsilon \nu^5\right]^{\frac{1}{4}}, \qquad (55)$$

It can be see that

$$y = y_0 \cdot [R_\lambda]^{\frac{3}{2}} \cdot z^{\sigma - \frac{1}{2}} \cdot e^{-z} \cdot U\left(2, \sigma + \frac{1}{2}, z\right), \qquad (56)$$

Where

$$y_0 = 2 \cdot \left[\frac{2}{3}\right]^{\frac{1}{4}} \cdot \sqrt{\frac{2\sigma}{\pi}} \cdot \frac{\Gamma\left(\frac{5}{2}\right)}{\Gamma(\sigma)} \cdot 10^{-\frac{3}{4}}, \qquad (57)$$

Before concluding this section it should be noted that the one-dimensional energy spectra based on the Kolmogorov scales could be derived entirely from the following set of steps:

[1] Give the calculation conditions: $\sigma, R_\lambda$;

[2] Calculating the variables: $x = \eta k$, $10^{-5} < x < 10^1$;

[3] Calculating the function $z$:

$$z = 2\sigma \cdot \left[\frac{3}{20}\right]^{\frac{1}{2}} \cdot [R_\lambda] \cdot [x]^2$$

[4] Calculating function $y$:

$$y \equiv F_{11}(k,t) \cdot [\varepsilon v^5]^{-\frac{1}{4}}$$
$$= y_0 \cdot [R_\lambda]^{\frac{3}{2}} \cdot z^{\sigma-\frac{1}{2}} \cdot e^{-z} \cdot U\left(2, \sigma + \frac{1}{2}, z\right)$$

Where

$$y_0 = 2 \cdot \left[\frac{2}{3}\right]^{\frac{1}{4}} \cdot \sqrt{\frac{2\sigma}{\pi}} \cdot \frac{\Gamma\left(\frac{5}{2}\right)}{\Gamma(\sigma)} \cdot 10^{-\frac{3}{4}}$$

[5] Plots of the logarithm of spectrum versus the logarithm of wave number: $\log y - \log x$

The implications of these slight differences in the length scale variation on the spectral scaling are summarized in Table 1.

**Table 1 Dependence of spectral scaling for one dimensional spectrum**

| Scaling | Wave number | Parameters | Spectrum |
|---|---|---|---|
| $L$ | $Lk$ | $\sigma$ | $F_{11} \cdot [bL]^{-1}$ |
| $\lambda$ | $\lambda k$ | $\sigma$ | $F_{11} \cdot [b\lambda]^{-1}$ |
| $\eta$ | $\eta k$ | $\sigma, R_\lambda$ | $F_{11} \cdot [\varepsilon v^5]^{-\frac{1}{4}}$ |
| $l$ | $lk$ | $a_1, \sigma$ | $F_{11} \cdot [bl]^{-1}$ |

# 5 Calculation results

A significant advantage of above analytical results is that they can be used to estimate one-dimensional energy spectra and make comparising. The appearance of aribitary additive constant might seem to be a drawback in the present theory. Actually, it is a necessity, in checking the theory by comparing with directly experiments. Based on the report in Ref. [14], we note that:

$\sigma = 0.8$ might be a good candidate for the physical solution. It is seen from the experiments

that the curves of $f\left(\frac{r}{\lambda}\right), b(t),$ and $\lambda(t)$ are in highly satisfactory agreement with theoretical results. Here we use this results directly. The low Reynolds number experimental data to be considered from G.Comte-Bellot and S.Corrsin [19] were obtained in wind tunnel. Figure 1 show plots of the logarithm of spectrum versus the logarithm of wave number using the Kolmogorov scales. Four different $R_\lambda$ for experimental data to be considered, and $R_\lambda$ are $36.6, 38.1, 41.1, 48.6$. (Table.2-b in Ref.19) The one-dimensional spectra measured from single probe singals at $\frac{U_0 t}{M} = 385, 240, 120, 45$. In this figure, they present, for four cases at $M = 2.54$, the one-dimensional energy spectra normalized with the Kolmogorov scales. In this way the effect of the dissipation, that dominates the energy spectra for such low $R_\lambda$, is enhanced.

The data scatter observed for high wave-numbers. The theoretical results are in agreement with the experiments, especially for the Reynolds effects. But it should be noted that: It is still unclear which is the best choice of the parameter $\sigma$ among various values. Quantitative information on the further properties of turbulent flow for higher $R_\lambda$, could also be obtained by use Eq. (56), with the one-dimensional energy spectra normalized with the Kolmogorov scales.

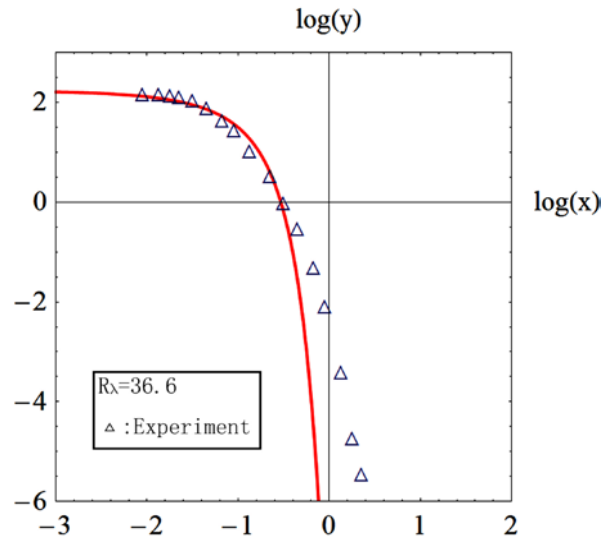

(a) $R_\lambda = 36.6, \frac{U_0 t}{M} = 385$

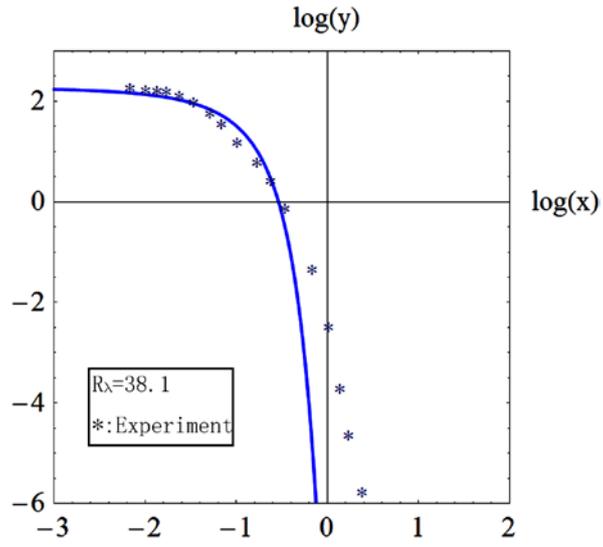

(b) $R_\lambda = 38.1, \dfrac{U_0 t}{M} = 240$

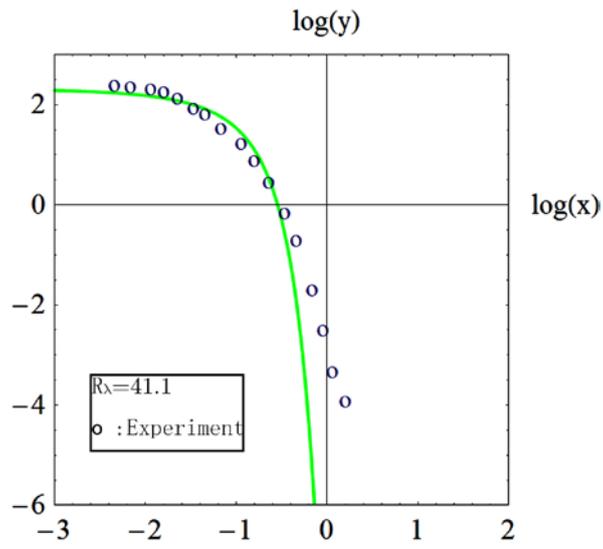

(c) $R_\lambda = 41.1, \dfrac{U_0 t}{M} = 120$

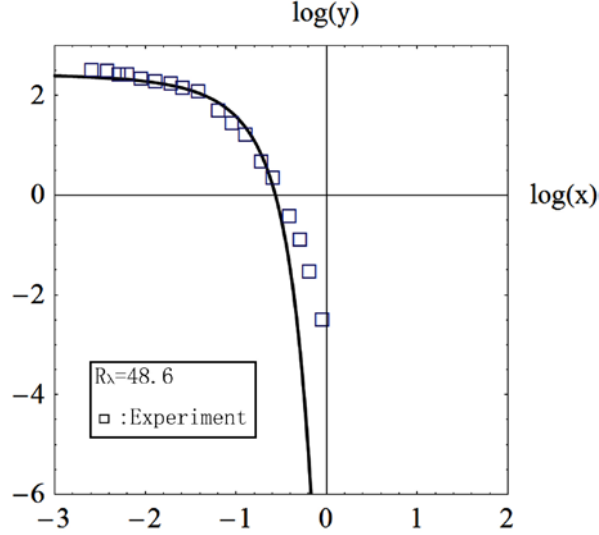

(d) $R_\lambda = 48.6, \dfrac{U_0 t}{M} = 45$

**Figure 1** The comparison for one-dimensional energy spectra at different $R_\lambda$

Finally we check the one-dimensional energy spectra and different parameters and find almost identical results as that derived in Eq. (56). This is reasonable because the basic parameters of one-dimensional spectra is $\sigma$, which produce the main part of the one-dimensional energy spectra, and lead almost unchanged in the form of one-dimensional energy spectra.

# 6 Conclusions

In summary we use the analytical method to explain the features of the isotropic turbulence one-dimensional energy spectra, including the rescaling issues and the related topic. Results show that the one-dimensional energy spectra presented in this paper agree well with the observations. As an additional result, our model can provide a natural explanation of the one-dimensional surrogate for the dimensionless of turbulent energy dissipation.

Furthermore we would like to discuss some implications of the model parameters. We note that only one single set of parameters is enough to explain the data. It may indicate that the sources are "standard" which have different parameters will have some effects on the observed fine structures of one-dimensional energy spectra. This work can be regarded as one progress approaching.

*This work was supported by the National Natural Sciences Foundation of China (Grant Nos. 11172162 and 90816013).*

## Appendix A. Gross properties of turbulence at various stations behind 2in. and 1.in grids (Table.4 in Ref.19)

| $M$ (cm) | $\dfrac{U_0 t}{M}$ | $\sqrt{\overline{u_1^2}}$ (cm sec$^{-1}$) | $\epsilon$ Dissipation rate (cm$^2$ sec$^{-3}$) | $\eta$ Kolmogorov microscale (cm) | $\lambda$ Taylor transverse microscale (cm) | $L$ transverse integral scale (cm) | $L_f$ longitudinal integral scale (cm) | $\dfrac{R_\lambda}{\sqrt{u_1^2}\lambda/\nu}$ | $\dfrac{\lambda}{L} R_\lambda$ |
|---|---|---|---|---|---|---|---|---|---|
| 5·08 | 42 | 22·2 | 4740 | 0·029 | 0·484 | 1·27 | 2·40 | 71·6 | 27·3 |
|  | 98 | 12·8 | 633 | 0·048 | 0·764 | 1·88 | 3·45 | 65·3 | 26·5 |
|  | 171 | 8·95 | 174 | 0·066 | 1·02 | 2·28 | 4·90 | 60·7 | 27·1 |
| 2·54 | 45 | 20·5 | 7540 | 0·026 | 0·355 | 0·60 | — | 48·6 | 28·7 |
|  | 120 | 10·6 | 731 | 0·046 | 0·581 | 0·90 | — | 41·1 | 26·5 |
|  | 240 | 6·75 | 145 | 0·069 | 0·845 | 1·07 | — | 38·1 | 30·0 |
|  | 385 | 5·03 | 48·5 | 0·091 | 1·09 | 1·20 | — | 36·6 | 33·2 |

TABLE 4. Gross properties of turbulence at various stations behind 2 in. and 1 in. grids

## Appendix B. Numerical data for one-dimensional spectra behind grids

$$(a)\ R_\lambda = 36.6, \frac{U_0 t}{M} = 385$$

| $k$ | $x$ | Experimental results | Theoretical results | error | Relative error |
|---|---|---|---|---|---|
| 0.1 | 0.0091 | 146.378 | 131.891 | -14.4868 | 0.1 |
| 0.15 | 0.01365 | 148.004 | 120.827 | -27.1776 | 0.18 |
| 0.2 | 0.0182 | 138.652 | 111.414 | -27.2381 | 0.2 |
| 0.25 | 0.02275 | 129.3 | 103.14 | -26.1606 | 0.2 |
| 0.35 | 0.03185 | 109.783 | 89.0105 | -20.7729 | 0.19 |
| 0.5 | 0.0455 | 77.6616 | 72.0151 | -5.64648 | 0.07 |
| 0.75 | 0.06825 | 43.9134 | 51.0105 | 7.09714 | 0.16 |
| 1 | 0.091 | 28.0558 | 36.1098 | 8.05409 | 0.29 |
| 1.5 | 0.1365 | 10.5717 | 17.6903 | 7.1186 | 0.67 |
| 2.5 | 0.2275 | 3.33416 | 3.66419 | 0.330032 | 0.1 |
| 3.5 | 0.3185 | 0.975852 | 0.593228 | -0.38262 | 0.39 |
| 5 | 0.455 | 0.292756 | 0.022733 | -0.27002 | 0.92 |
| 7.5 | 0.6825 | 0.048386 | 2.08E-05 | -0.04837 | 1 |
| 10 | 0.91 | 0.00801 | 2.35E-09 | -0.00801 | 1 |
| 15 | 1.365 | 0.000388 | 7.78E-12 | -0.00039 | 1 |
| 20 | 1.82 | 1.84E-05 | 1.38E-11 | -1.8E-05 | 1 |
| 25 | 2.275 | -3.5E-06 | 1.08E-11 | -3.5E-06 | 1 |

(d) $R_\lambda = 48.6, \dfrac{U_0 t}{M} = 45$

| $k$ | $x$ | Experimental results | Theoretical results | error | Relative Error |
|---|---|---|---|---|---|
| 0.1 | 0.0026 | 329.489 | 232.504 | -96.9846 | 0.29 |
| 0.15 | 0.0039 | 315.664 | 223.506 | -92.1578 | 0.29 |
| 0.2 | 0.0052 | 274.19 | 215.701 | -58.4886 | 0.21 |
| 0.25 | 0.0065 | 266.126 | 208.693 | -57.4325 | 0.22 |
| 0.35 | 0.0091 | 222.347 | 196.311 | -26.0364 | 0.12 |
| 0.5 | 0.013 | 195.85 | 180.456 | -15.3943 | 0.08 |
| 0.75 | 0.0195 | 173.961 | 158.555 | -15.4054 | 0.09 |
| 1 | 0.026 | 144.007 | 140.351 | -3.65589 | 0.03 |
| 1.5 | 0.039 | 122.118 | 111.14 | -10.9779 | 0.09 |
| 2.5 | 0.065 | 50.6906 | 70.5835 | 19.8929 | 0.39 |
| 3.5 | 0.091 | 27.8798 | 44.6331 | 16.7532 | 0.6 |
| 5 | 0.13 | 16.9353 | 21.7949 | 4.8596 | 0.29 |
| 7.5 | 0.195 | 4.74648 | 5.9049 | 1.15842 | 0.24 |
| 10 | 0.26 | 2.23499 | 1.35438 | -0.88061 | 0.39 |
| 15 | 0.39 | 0.382484 | 0.040288 | -0.3422 | 0.89 |
| 20 | 0.52 | 0.131335 | 0.000522 | -0.13081 | 1 |
| 25 | 0.65 | 0.030414 | 2.78E-06 | -0.03041 | 1 |
| 35 | 0.91 | 0.003237 | 1.01E-11 | -0.00324 | 1 |